# Swift Heavy Ion Induced Modification Studies of $C_{60}$ Thin Films

Navdeep Bajwa[1], Alka Ingale[2], D.K. Avasthi[3], Ravi Kumar[3],
K. Dharamvir[1] and V.K. Jindal[1]

[1]*Department of Physics, Panjab University, Chandigarh-160014, India*
[2]*Laser Physics Division, Center For Advanced Technology, Indore-452013 India*
[3]*Nuclear Science Center, Aruna Asaf Ali Marg, New Delhi-160067, India*

Modification induced by 110 MeV Ni ion irradiated thin film samples of $C_{60}$ on Si and quartz substrates were studied at various fluences. The pristine and irradiated samples were investigated using Raman spectroscopy, electrical conductivity and optical absorption spectroscopy. The Raman data and band gap measurements indicate that swift ions at low fluences result in formations that involve multiple molecular units like dimer or polymer. High fluence irradiation resulted in sub-molecular formations and amorphous semiconducting carbon, indicating overall damage of the fullerene molecules. These sub-molecular units have been identified with nanocrystalline diamond and nanocrystalline graphite like formations.

## INTRODUCTION

Ever since its synthesization in the laboratory[1], the fullerene solid has been a subject of significant interest as target for ion–beam irradiation. The interest was stimulated by the potential of fullerene material towards superconductivity by implanting foreign ions in the fullerene cage and other applications like optical limiters, electrical storage devices, $C_{60}$ based diodes and transistors etc. Irradiation effects of keV and some MeV energies has been investigated[2-14]. In general, it has been observed by various research groups that the $C_{60}$ materials undergo heavy damage by low energy ion irradiation in the energy range 30 - 300 keV of various ions [2-6]. There has also been study on irradiation effects of MeV energies on fullerene films[7-9, 12-14]. Itoh et al[7] have reported fragmentation of $C_{60}$ induced by impacts of 2 MeV $Si^{4+}$ ions. T.Le.Brun et al[8] have reported ionization and fragmentation in $C_{60}$ molecules by impacts of Xe ions in the energy range of 420 – 625 MeV. Subsequently[9], there has been some interest in studying the phase transformations of the $C_{60}$ solids using MeV energies. It has been reported that under suitable conditions of temperature and pressure, the $C_{60}$ undergoes a dimer or a polymeric phase transformation[10, 11]. Ion beam irradiation could provide similar conditions so that the formation of aggregates (of molecules of $C_{60}$) is favoured. This could further result in the formation of solids of these aggregates. The energy of the ions and their fluence play a key role in the end product of the target.

Energetic ions loose energy during their passage in a material. The strength of interaction of the incident ion in a material i.e., with the electrons of the atoms in the target material (including the core electrons of the constituent atoms or molecules) as well as with the nuclei (which happen to be the mass centers) depends on the mass, charge and energy of the incident ion. Therefore, it is of interest to study the interaction of ion beams with fullerene solids using swift ions at various fluences. The interaction of the incident ion with the atomic radii (or molecular mass center) is an elastic collision process and is prominent at low energy (eV to few hundred keV). On the other hand the interaction of the incident ion with the electrons in the material is an inelastic process and is prominent at higher energies (a few tens of million electron volts and higher). The energy loss of ion beams in the former case (of interaction with the nuclei) is termed as nuclear stopping, designated by $S_n$, and for the latter case is designated as $S_e$.

In this paper an attempt has been made to report the changes in the structural and transport profiles of $C_{60}$ by 110 MeV, Ni ion irradiation at different fluences. This would provide a controlled study of what is formed due to the dominant electronic energy loss (i.e., ion interactions prominent at few tens of MeV and higher energies) at various fluence ranges.

## EXPERIMENTAL DETAILS

Thin films of $C_{60}$ were deposited on various substrates such as Si(100) and quartz using resistive heating method. Commercially available powder of $C_{60}$, with a purity of 99.9%, was palletized and used for film deposition. Deposition was performed in a high vacuum ($2x10^{-6}$ Torr) environment by sublimation of the pellets at a rate of ~ 0.1nm/s and by passing a current of ~ 75A in a Ta boat. The thickness of the film was measured using a quartz crystal oscillator. Films of thickness 160 nm and 230 nm were deposited on Si and quartz substrates respectively. The thickness of the $C_{60}$ film under study was kept very small as compared to the range (~19 μm) of the Ni ion used. Using the code TRIM 95[15], $S_e$ and $S_n$ values for $^{58}Ni$ ion of 110MeV energy were calculated. The $S_e$ was obtained to be $7.28 \times 10^2$ eV/Å and $S_n$ value was obtained to be $9.89x10^{-1}$ eV/Å respectively. The $S_e$ value being about 3 orders of magnitude more than the $S_n$ value shows that the electronic energy loss is dominant for the irradiation of $^{58}Ni$ ion at 110MeV. The thin films of $C_{60}$ were characterized using Raman spectroscopy and Optical Absorption measurements. Raman spectra of the thin films were measured using Raman spectrometer, Renishaw, Model 1000. Micro-Raman data was recorded at room temperature with Ar ion laser excitation 5145Å (~ 2mW with 50X objective) at Indian Diamond Institute, Surat, Gujrat in a range of 100-2500cm$^{-1}$. Optical absorption measurements were performed using U3300 HITACHI spectrophotometer at NSC. The well-characterized samples were irradiated with 110MeV $^{58}Ni$ ions at a fluence ranging from $10^{11}$ - $10^{14}$ ions/cm$^2$ using 15UD Pelletron Accelerator at NSC, New Delhi. Studies by Raman and Optical absorption spectroscopy, In-situ and low temperature conductivity were made on the irradiated $C_{60}$ films in order to understand the effects of irradiation.

## RESULTS AND DISCUSSION:

### A. Raman Spectroscopy

Raman spectra of unirradiated and irradiated $C_{60}$ with various fluences for Ni ion after appropriate calibration and baseline fitting of the raw data are shown in Fig. 1. Various Raman active modes of $C_{60}$ molecule as well as that in bulk were observed. The most intense mode at 1465 $cm^{-1}$ corresponds to the pentagon pinch mode[16] (the $A_g$ internal mode of $C_{60}$). It was observed that the intensity of all the $C_{60}$ modes decreases with increase in fluence of Ni ions (Fig. 1). Irradiation effects show that there are additional features arising from internal modes close to the pentagon pinch mode ($A_g$ internal mode of $C_{60}$) prominently existing in most of the Raman measurements as shown in Fig. 1. A plot of these features around the pentagon pinch mode has been shown in Fig. 2, where emphasis has been given to softer modes close to this mode, for $C_{60}$ samples irradiated with Ni on Si substrate. The results indicate that the intensity of the dominant $C_{60}$ mode (1465 $cm^{-1}$) diminishes as the fluence of incident ions is increased, with ever increasing asymmetry of this mode towards lower wave numbers. When this asymmetrical line profile is fitted to two Lorentzians, the broad asymmetrical peak is nicely represented as a mixture of two peaks, i.e. in addition to the mode of $C_{60}$ at 1465 $cm^{-1}$, a new peak corresponding to a relatively soft mode around 1458 $cm^{-1}$, also starts to build. This peak has been attributed to a polymer mode, and rises in intensity with increasing ion fluences. This mode has already been observed in photoluminescence experiments[17,18]. It has been reported that the soft mode at 1458 $cm^{-1}$ is perturbed by the formation of dimers, trimers and higher oligomers of $C_{60}$. Raman results of present work (i.e.,110 MeV Ni ion irradiation on $C_{60}$ film), indicate that this peak starts to develop at fluences of the smallest incremental value and grows to a maximum at low fluences and then it decreases with fluence. These results for polymer formation seem to be due to electronic energy loss component of the incident ions. The plot in Fig. 3 shows the content of $C_{60}$ in the thin film as a function of fluence which is a measure of the amount of damage that takes place with increase in fluence and the corresponding rise and decay of the polymer peak with increase in fluence. The Raman measurement for high fluence (i.e. after a fluence of $2x10^{13}$ ions/$cm^2$) is shown in Fig.4. It indicates that in and around the $A_g$ mode there is a further drift from the above stated behaviour in the phase transformation of $C_{60}$. The amorphized $C_{60}$ shows a broad peak around 1360 $cm^{-1}$ (which can be attributed to the disordered peak of graphite, also known as the D peak of graphite) along with peaks around 1582$cm^{-1}$ and 1220$cm^{-1}$. There is no $C_{60}$ peak at 1465 $cm^{-1}$ indicating its complete destruction. The peak around 1582 $cm^{-1}$ is the signature of graphite[19] also referred to as the G peak (G for graphite) and its being broad indicates some deviation from the bulk crystalline graphite. It could be possibly due to micro- or nano- graphite-like formations[20-22]. The Raman results indicate that this peak starts to emerge even at lower fluences, where it could not be resolved due to the presence of a $C_{60}$ mode near it. Only after the complete destruction of $C_{60}$ this peak becomes visible at the position identified by graphite. The peak around 1360$cm^{-1}$ can be attributed to first order scattering from zone boundary phonon activated by disorder, also referred to as the D peak (D for disorder) and the peak around 1220$cm^{-1}$ indicates that there is also some formation of nano crystalline diamond[20]. Additional structure around 1453$cm^{-1}$ is suddenly found to appear at this high fluence, which cannot be attributed to any known formations.

This could be a transitory mode of mixed state of polymer and amorphous carbon. Further experimental study at higher fluences is needed to identify it.

When an energetic ion passes through a solid film it gives rise to a localized zone of high temperature and pressure as shown in Fig. 5. There is a radial profile of temperature surrounding the ion path. The maximum damaged part, is a cylindrical zone of radius R where the temperature exceeds the melting temperature of the material for a duration of the order of picosecond. The extent of damage and the cross-section of this damaged zone depends on the electronic energy loss $S_e$. Further away from this core zone, the solid is likely to undergo structural phase transformations (defects in the crystal structure as well as polymerization, i.e. joining by chemical bond, of two or more buckyballs) due to the heat flow, which spreads away from the center of the zone. It is expected that maximum damage of $C_{60}$ takes place within a certain cylindrical diameter around the ion path and the $C_{60}$ in the surrounding region transforms to some dimer/polymer material. Now the Raman results obtained above and also shown in Fig. 2 to 4 can be explained using the above concept. At low fluences $C_{60}$ gets damaged along the ion path and amorphization starts with the onset of the first fluence. A damaged zone formed radially along the ion path has high temperature and pressure and results in amorphization. Heat flow around the damaged zone results in structural transformation around the damaged zone like dimerization/polymerization. The structural changes are maximum at low fluences where the damage due to the ion irradiation does not destroy the entire $C_{60}$ content present in the thin film and there is ample probability for dimer/polymer features to form around the damaged zone. With the increase in fluence, damage of $C_{60}$ as well as the phase transformed features like dimer/polymer, takes place and at high fluences there is a complete overlap of damage zones with no $C_{60}$ or any polymerized $C_{60}$ phase left in the thin film. This possibly explains the above observed features in the Raman spectra of Ni ion irradiated $C_{60}$ thin films.

## B. Optical Absorption Spectroscopy

Optical Absorption studies were performed on 110 MeV Ni ion irradiated samples of $C_{60}$ with quartz substrate in the range of 200 – 800 nm wavelength. The spectral intensity of the light transmitted through the specimen determines the absorption edge, which is a measure of the bandgap in a non-conducting solid. Bandgap determination was performed in the absorption mode with light of UV-VIS range. The transmitted light through the sample was made incident on the detector (photo multiplier tube) and the data was recorded in computer subtracting the reference signal from the sample signal.

Fig. 6 shows the absorption spectra of the pristine as well as Ni irradiated $C_{60}$ at certain fluences. Only those measurements that were made on quartz substrate have been useful for this study because quartz does not interfere in the UV-VIS range of the optical spectra. Using the absorption data, band gaps for various irradiated samples were determined. It was observed that with the increase in fluence, the absorption edge changes and hence the band gap. A plot for band gap variation with fluence for Ni irradiated $C_{60}$ is shown in Fig. 7. It has been shown in literature that for pristine $C_{60}$ thin film, the size of the band gap varies progressively with the degree of crystallinity of the fullerene film [23-29]: highly crystalline films have $E_g$ values close to 1.5eV[23] and highly disordered films have band gap values close to 2.3eV[29]. The present work shows irradiation effect of $C_{60}$ thin film with the pristine $C_{60}$ film having a band gap value of 1.97eV.

The optical absorption data presented in Fig. 6 and 7, were observed to be in conformity with the Raman measurements. As the ion fluence was increased, it was observed that above a fluence of $1x10^{13}$ ions/cm$^2$, the absorption bands in the absorption spectra of pristine $C_{60}$ completely disappear and so does the $C_{60}$ peak in the Raman spectra. The band gap data also shows consistency with the results obtained from Raman spectroscopy. The initial high band gap behaviour changes into very low band gap behaviour at high fluences. The high band gap can be attributed to the $C_{60}$ (which remains undestroyed in the thin film at low fluences) as well as some $C_{60}$ polymer material, which gets formed at low fluences. The change to low band gap at higher fluences appears to be a result of some amorphous semi-conducting and nanocrystalline like nano-/micro- crystalline graphite formations.

## C. Conductivity measurements

Conductivity measurements were performed on pristine and irradiated sample of $C_{60}$ in vacuum at various steps. In-situ measurements were performed for Ni ion irradiated samples on quartz substrate. The result of in-situ resistivity for $C_{60}$ sample irradiated at various fluences of Ni ions, is given in Fig. 8. At low fluences, the resistivity was high and two-probe method was used to determine the conductivity of the sample. As the fluence was increased the conductivity of the irradiated $C_{60}$ sample was observed to increase and four-probe method was used to determine the conductivity.

It was observed that samples exposed to fluences higher than $10^{13}$ ions/cm$^2$ showed a significant increase in conductivity. Therefore, to determine the nature of the transformation in the material, temperature dependent resistivity measurements were performed using four-probe method. For such highly conducting samples the band gap was found to be too small and could not be determined by optical absorption spectroscopy. The resistance R of a semiconducting sample of bandgap $E_g$ varies with temperature as

$$R = R_0 \exp(-E_g / 2k_B T)$$

Accordingly, a plot of $\ln(R/R_0)$ vs. 1/T (called Arrhenius plot) determines the bandgap. The bandgap values of highly conducting Ni ion irradiated samples were obtained using the Arrhenius plot as shown in Fig. 9. The plot in Fig. 7 shows these band gap values as a function of fluence, supplementing the values determined using optical absorption data.

**SUMMARY AND CONCLUSION:**

This paper attempts to report the effects of swift heavy ions on thin films of $C_{60}$. The role of low energy ions has already been studied extensively. The investigations using Raman spectroscopy and Optical Absorption spectroscopy supported by conductivity measurements reveal that at low fluences there is formation of multiple molecular units like dimers and polymers. High fluences result in submolecular formations and amorphous semiconducting carbon. This finds possible explanation from our understanding that when an energetic ion passes through the thin film of $C_{60}$ it creates a cylindrical zone containing amorphized /damaged $C_{60}$, surrounded by a region containing polymerized fullerene. The polymer content increases with fluence up to a certain value and then decreases, disappearing at higher fluences. At higher fluences, complete overlap of these zones produce total damage of $C_{60}$, resulting in amorphous low band gap carbon containing nano-/micro- crystalline graphite. Band gap is observed to be changing from 1.97eV for pristine to 0.12eV at a fluence $10^{14}$ ions/cm$^2$. This also explains the increased conductivity at higher fluences.

The present work leads to the conclusion that swift heavy ions, due to their role in losing most of the energy to electronic component of the target, transform the structure of $C_{60}$. Therefore, swift heavy ions of moderate fluences ($<10^{12}$ ions/cm$^2$) would be ideally suited for aggregate formation in the form of dimer or polymer $C_{60}$. High fluences of swift heavy ions result in amorphization of $C_{60}$ where the formation of nanocrystalline graphite and nanocrystalline diamond is evident. New features observed at high fluences ($\sim 10^{14}$ ions/cm$^2$ ) motivate us to carry out further experiments at higher fluences.

**ACKNOWLEDGMENTS**

VKJ, KD and NB gratefully acknowledge financial support for the research project from the Department of Science and Technology, Government of India, grant number SP/S2/M13/96. KD and NB also acknowledge financial support from the Nuclear Science Center, New Delhi. We are also thankful to Ambuj Tripathi for some very useful discussions and Asiti Sharma for his invaluable help in studying the absorption spectroscopy.

**REFERENCES**

1. W. Kratschmer, L.D. Lamb, K. Fostiropouios, and D.R. Huffmann, Nature (London) **347**, (1990) 354. -ibid Chem.Phys.Lett. **17**,(1991) 167.
2. J.Kastner, H.Kuzmany, L.Palmetshofer, P.Bauer, G.Stingeder, NIMB **80/81** (1993) 1456.
3. Zhong-Min Ren , Xing-Long Xu, Yang-Cheng Du, Zhi-Feng Ying, Xia-Xing Xiong and Fu-Ming Li NIMB **100** (1995) 55.
4. C.E.Foerster, F.C.Serbena, C.M.Lepienski, D.L.Baptista, F.C.Zawislak, NIMB **148** (1999) 634.

5. F.C.Zawislak, D.L.Baptista, M.B.Behar, D.Fink, P.L.Grande, J.A.H.da Jornada NIMB **149** (1999) 336.
6. L.Palmetshofer and J.Kastner, NIMB **96** (1995) 343.
7. A.Itoh, H. Tsuchida, K.Miyabe, M. Imai, B.Imanishim NIMB **129** (1997) 363.
8. T.LeBrun, H.C.Berry, S.Cheng, R.W.Dunford, H.Esbensen, D.S.Gemmell and E.P.Kanter Phys.Rev.Letts. **72**,(1994) 3965.
9. D.K.Avasthi, Vacuum, **47**,(1996) 1249.
10. S.M.Bennigton et al. J. Phys. Condens. Matter **12** (2000) 51.
11. J.Winter and H.Kuzmany, Phys.Rev.**B88**(2000) 51.
12. S.Lotha, A.Ingale, D.K.Avasthi, V.K.Mittal, S.Mishra, K.C.Rustagi, A.Gupta, V.N.Kulkarni, D.T.Khathing, Solid State Communications **111**(1999) 55.
13. R.M.Papaleo, A.Hallen, J.Eriksson, G.Brinkmalm, P.Demirev, P.Hakansson and B.U.R. Sundqvist NIMB **91** (1994) 124.
14. Ch.Dufour, E.Paumier, M.Toulemonde, NIMB **122** (1997) 445.
15 J.F.Ziegler, J.P.Biersack, V.Littmark, The Stopping and Range of Ions in Solids, Pergamon Press, New York,1985.
16. Z.H.Dong, P.Zhou,J.M.Holden,P.C.Eklund,M.S.Dresselhaus and G.Dresselhaus, Phys. Rev.B, **48**, (1993) 2862.
17. M.S.Dresselhaus, G.Dresselhaus and P.C.Eklund, *Science of Fullerenes and Carbon Nanotubes*, Academic Press, 1996.
18. Y.Wang, J.M.Holder, Z.H.Dong, X.X.Bi and P.C.Eklund, Chem.Phys.Lett. **211** (1993) 341.
19. M.Tamor and W.Vassel, J.Appl. Phys. **76**, (1994) 3823.
20. J.Schwan, S.Ulrich, V.Batori, H.Ehrhardt and S.R.P.Silva, J.Appl.Phys. **80(1)** (1996) 440
21. F.Tunistra and J.Koenig, J.Chem.Phys.**53**, (1970) 1126.
22. S.Ghosh, A..Ingale, T.Som, D.Kabiraj, A.Tripathi, S.Mishra, S.Zhang, X.Hong, D.K.Avasthi, Solid State Communications **120** (2001) 445.
23. C.Wen, T.Aida,I.Honma,H.Komiyama, and K.Yamada, J.Phys:Condens.Matter **6**, (1994) 1603.
24. K.Kamaras, A.Breitschwerdt, S.Pekker, K.Fodor-Csorba, G.Faigel, and M.Tegze, Appl.Phys.A **56**, (1993) 231.
25. P.L.Hansen, P.J.Fallon, and W.Kratchmer, Chem. Phys.Lett. **181**, (1991) 367.
26. J.L.Sauvajol, Z.Hricha, N.Coustel, A.Zahab,and R.Aznar, J.Phys.: Condens.Matter **5**, (1993) 2045.
27. R.K.Krener, J.Rabenau, W.K.Maser, M.Kaiser, A.Simon, M.Haluska, and H.Kuzmany, Appl. Phys.A **56**, (1993) 211.
28. P.N.Saeta, B.L.Greene,A.R.Kortan, N.Kopylov, and F.A.Thiel, Chem. Phys. Lett. **190** (1992) 184.
29. R.W.Lof, M.A.van Veenendaal, B.Koopmans, H.T.Jonkman and G.A.Sawatzky, Phys. Rev. Lett. **68**, (1992) 3924.

**FIGURE CAPTION:**

Fig.1: Raman spectra of Pristine and Ni irradiated thin film samples of $C_{60}$ on Si substrate at various fluences.

Fig.2: Raman data measurements fitted around the pentagonal pinch mode of $C_{60}$ ($A_g$ mode.). Onset of polymerization around 1458cm$^{-1}$ is evident at low fluences of Ni ions for $C_{60}$ films on Si substrate.

Fig.3: Plot showing the content of $C_{60}$ in thin film with increase in fluence. Simultaneously, the growth and decay of polymerization with increase in fluence is also shown. The arrow at the lower left corner shows that there is no content of $C_{60}$ polymerization before irradiation in the pristine film. The arrow on the upper left corner shows the content of $C_{60}$ in the pristine $C_{60}$ film.

Fig.4: Plot showing amorphization of $C_{60}$ on Si substrate at high fluences for Ni ion irradiation at a fluence of $1x10^{14}$ ions/cm$^2$.

Fig.5: Model for the explanation of the experimental results obtained.

Fig.6: Plot showing absorption spectra of pristine and Ni ion irradiated thin film samples of $C_{60}$ on Quartz substrate at certain fluences.

Fig.7: Plot showing band gap variation with increase in fluence for Ni ion irradiated $C_{60}$ film on quartz substrate.

Fig.8: Plot showing resistivity versus fluence for Ni ion irradiated $C_{60}$ film on quartz substrate, showing onset of conductivity at a fluence of $10^{13}$ ions/cm$^2$.

Fig.9: Resistivity measurements on high fluence Ni irradiated $C_{60}$ samples as a function of temperature.

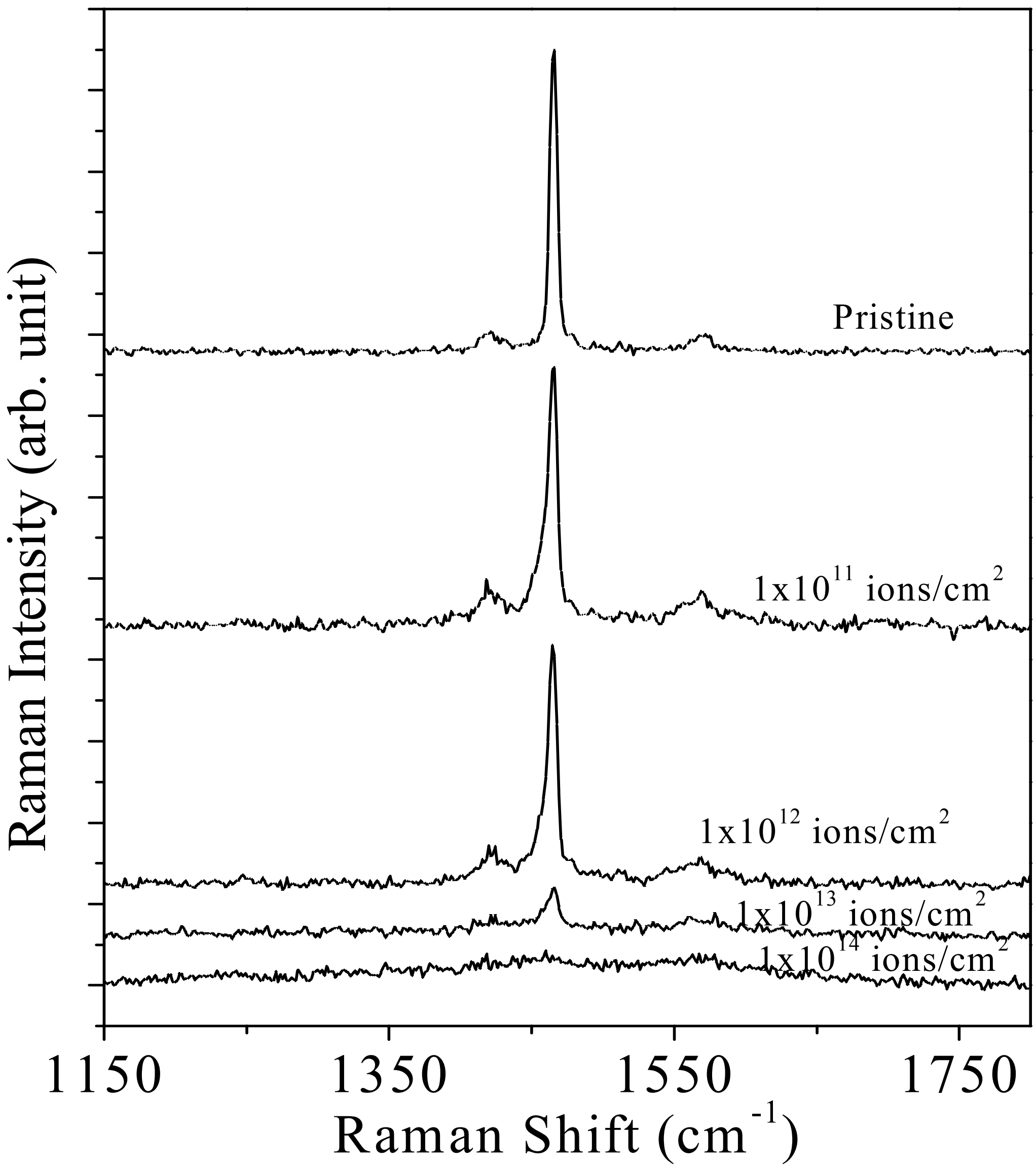

Fig. 1

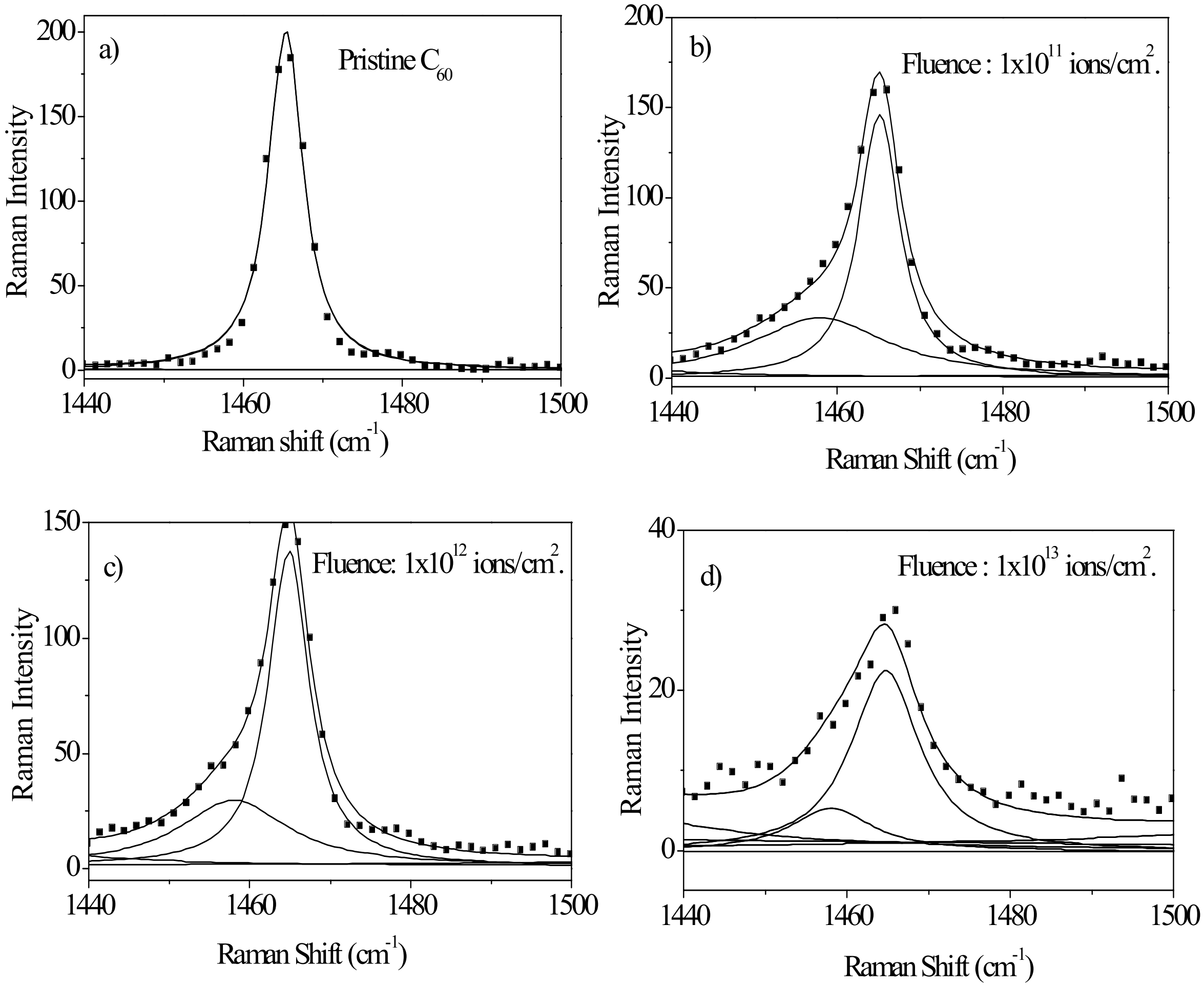

Fig. 2

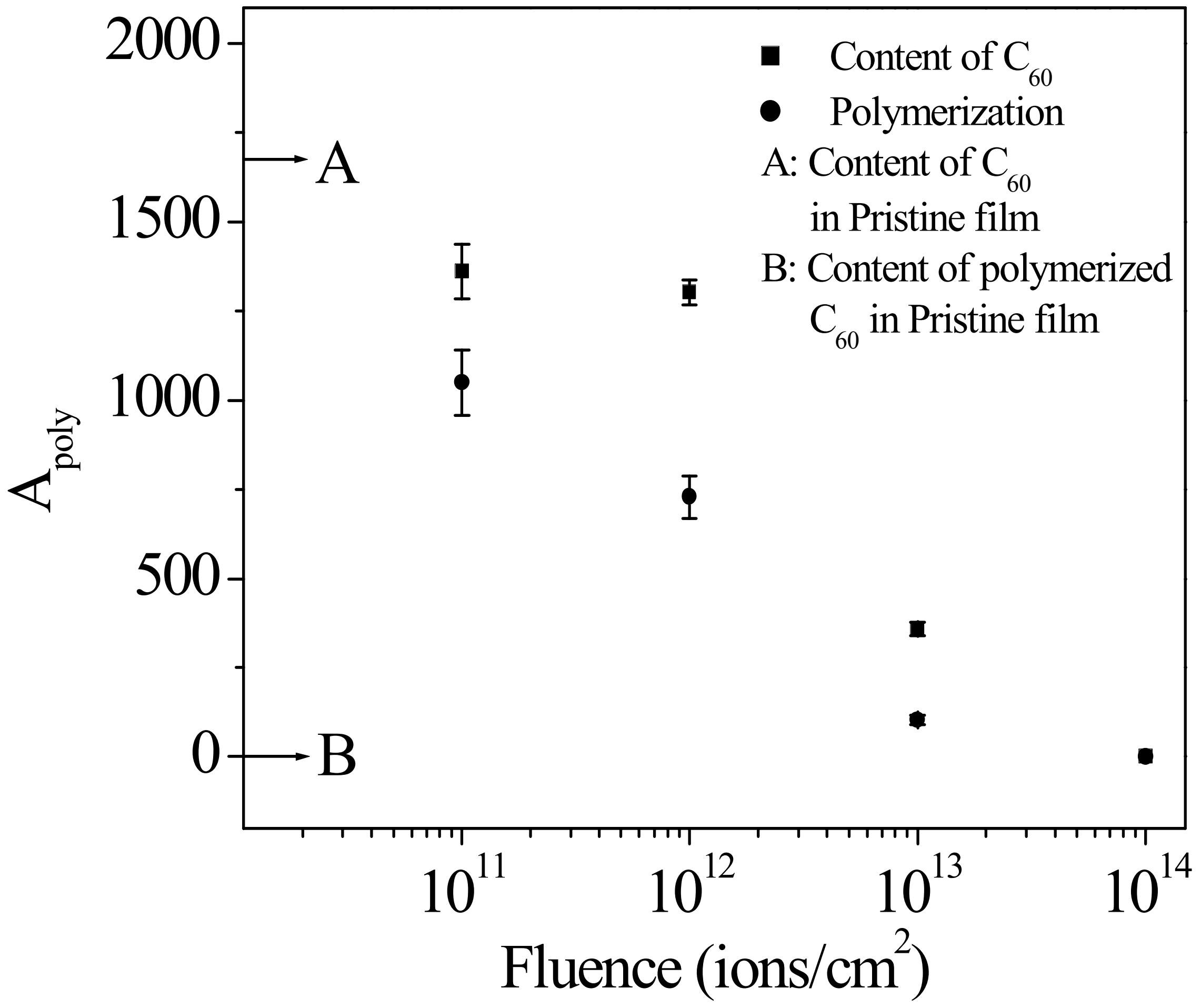
2000
1500
1000
500
0
A
B
A<sub>poly</sub>
Content of C60
Polymerization
A: Content of C60 in Pristine film
B: Content of polymerized C60 in Pristine film
10^11
10^12
10^13
10^14
Fluence (ions/cm^2)

Fig. 3

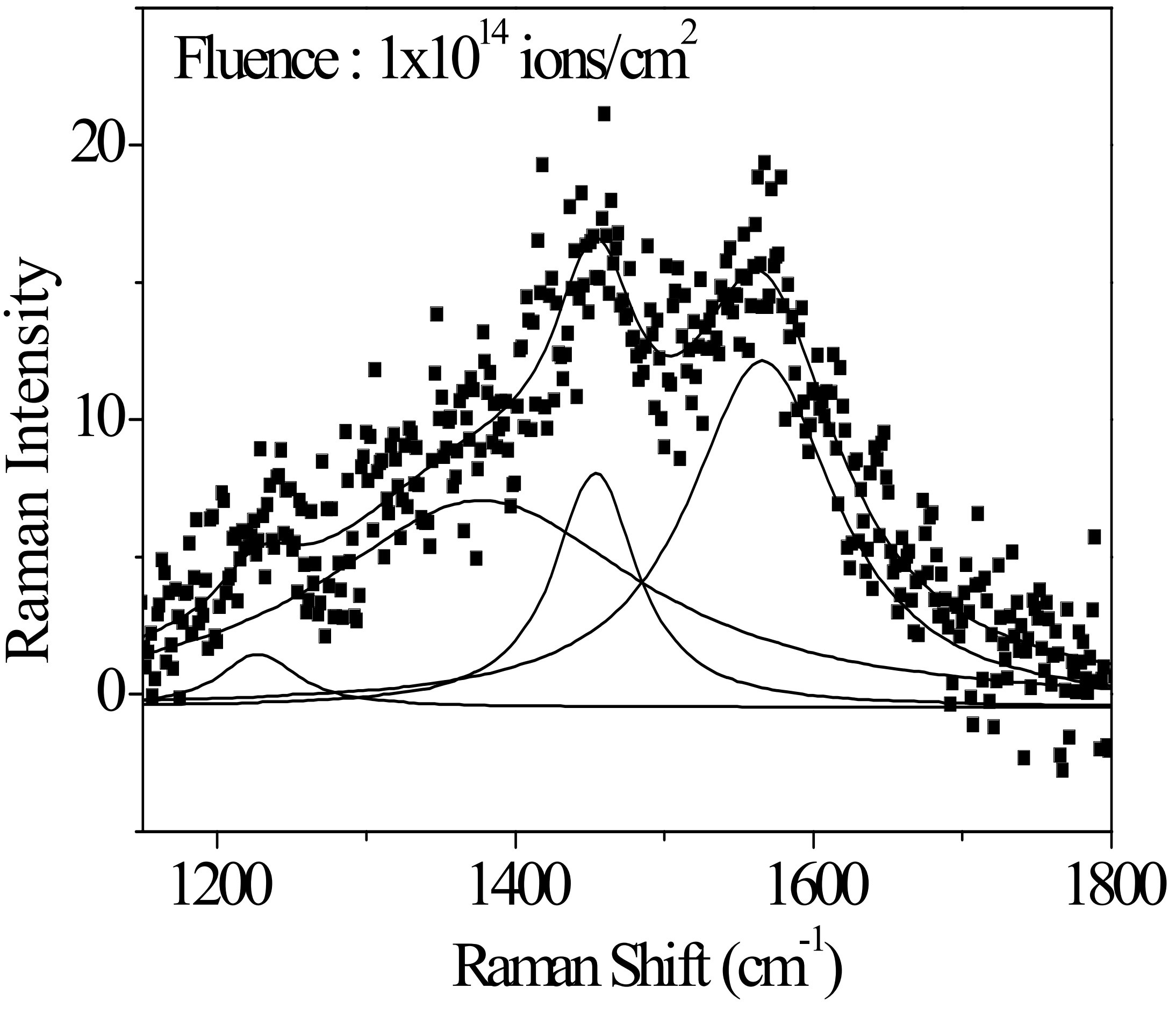

Fig. 4

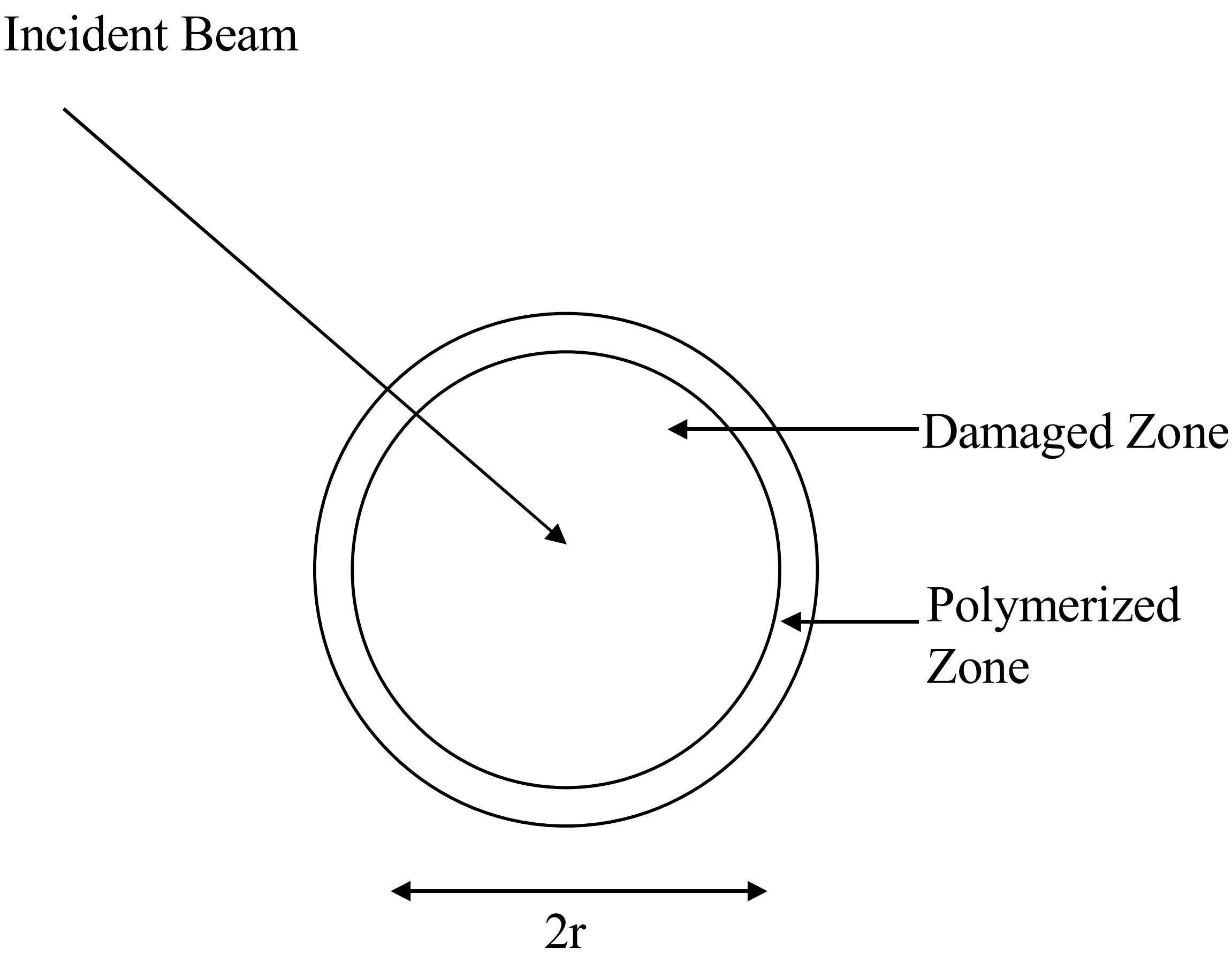

Fig. 5

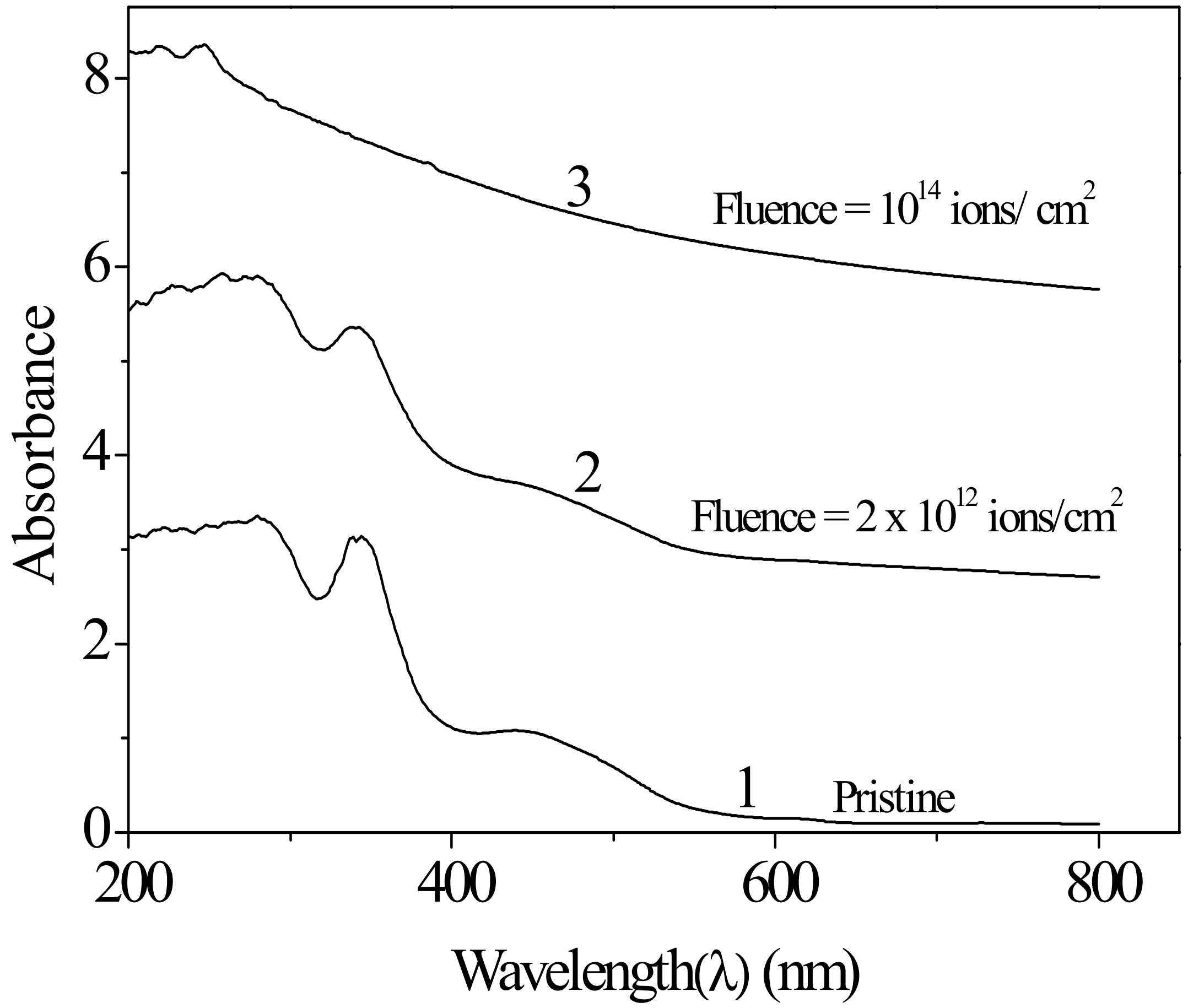

Fig. 6

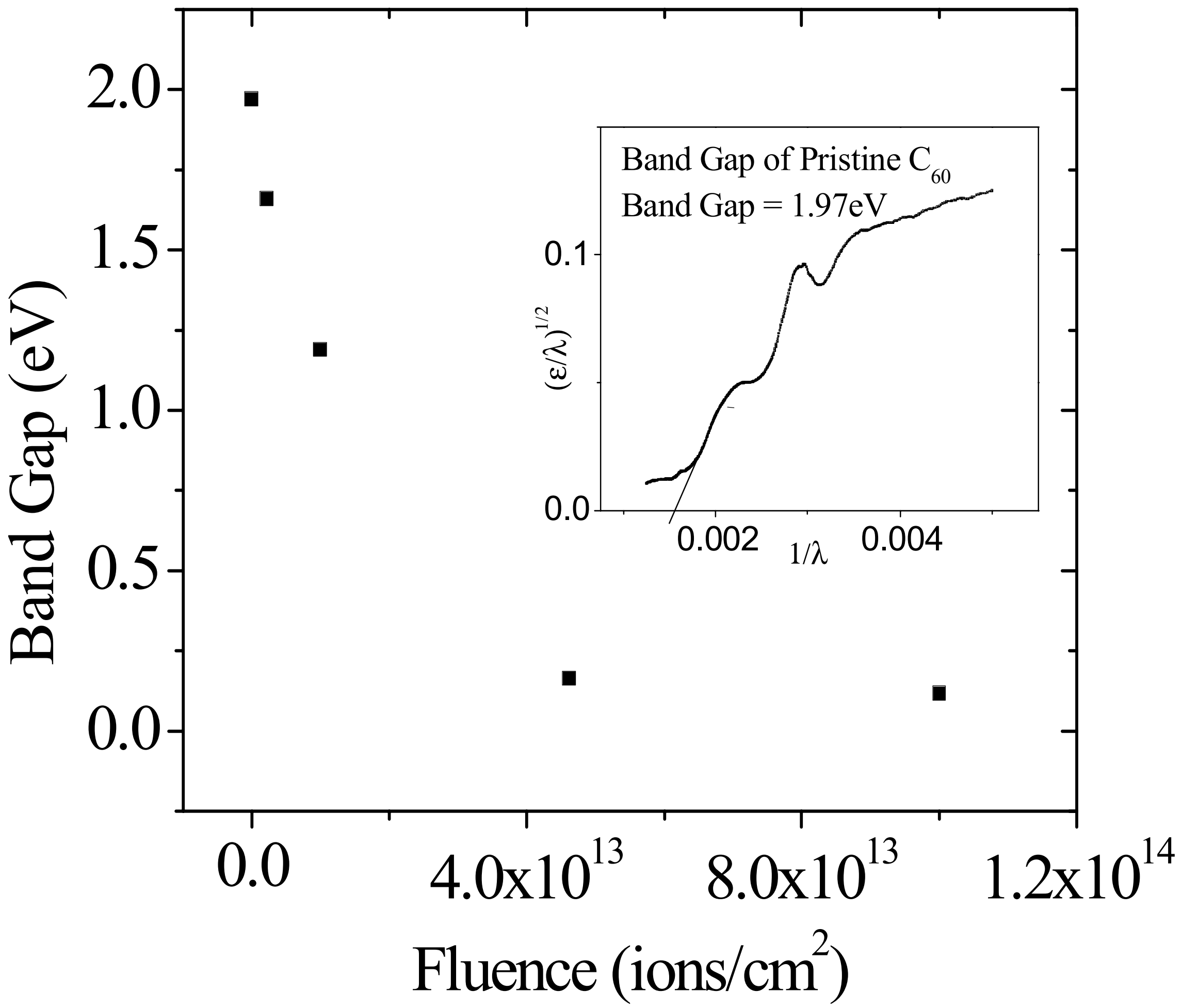

Fig. 7

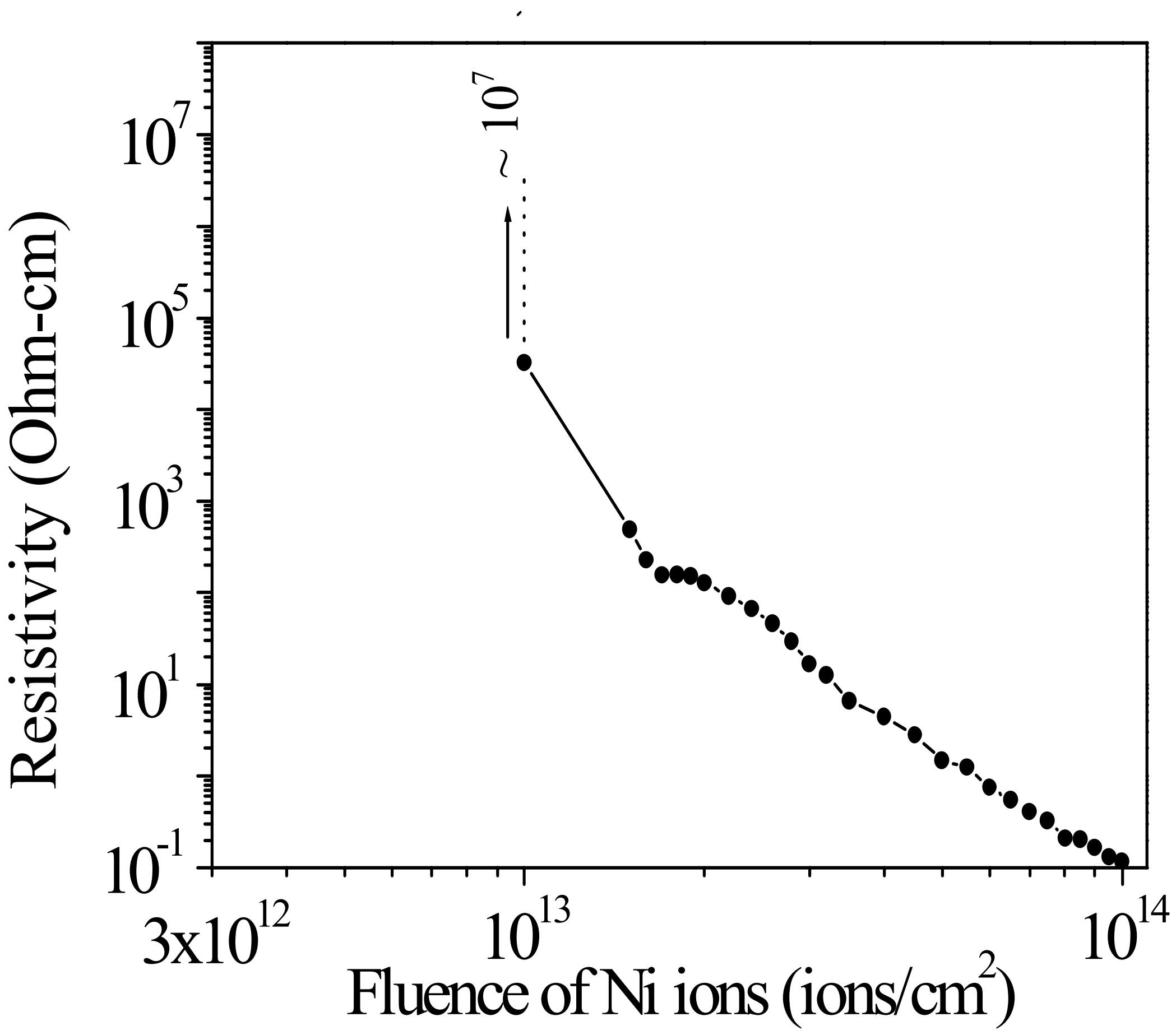

Fig. 8

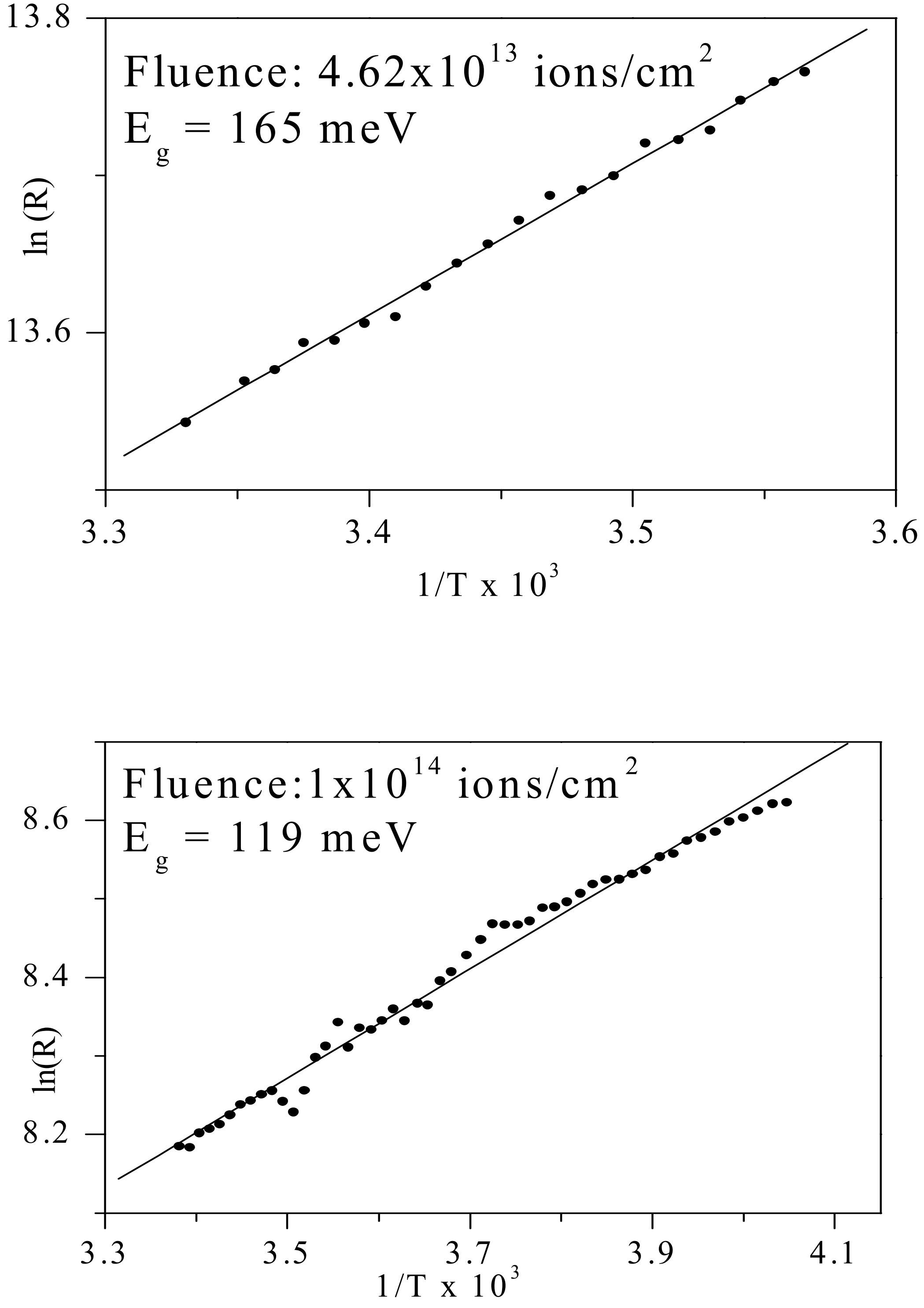

Fig. 9